\documentclass[runningheads]{llncs}
\usepackage[T1]{fontenc}
\usepackage{graphicx}
\usepackage{booktabs}
\usepackage{multirow}
\usepackage{hyperref}
\usepackage{fontawesome}
\usepackage{pifont}
\usepackage{subcaption}
\usepackage{todonotes}
\usepackage{wrapfig}
\newcommand{\cmark}{\ding{51}}%
\newcommand{\xmark}{\ding{55}}%
\usepackage{color}

\begin{document}
\title{CoRECT: A Framework for Evaluating Embedding Compression Techniques at Scale}
\titlerunning{CoRECT: Evaluating Embedding Compression Techniques at Scale}
\author{Laura Caspari\inst{1}\orcidID{0009-0002-6670-3211} \and
Michael Dinzinger\inst{1}\orcidID{0009-0003-1747-5643} \and
Kanishka Ghosh Dastidar\inst{1}\orcidID{0000-0003-4171-0597} \and
Christofer Fellicious\inst{1}\orcidID{0000-0001-7487-7110} \and
Jelena Mitrovi\'{c}\inst{1}\orcidID{0000-0003-3220-8749} \and
Michael Granitzer\inst{1, 2}\orcidID{0000-0003-3566-5507}}
\authorrunning{L. Caspari et al.}
\institute{University of Passau, Passau, Germany\\
\email{\{laura.caspari, michael.dinzinger\}@uni-passau.de} \and
Interdisciplinary Transformation University Austria, Linz, Austria}
\maketitle

\begin{abstract}
Dense retrieval systems have proven to be effective across various benchmarks, but require substantial memory to store large search indices.
Recent advances in embedding compression show that index sizes can be greatly reduced with minimal loss in ranking quality.
However, existing studies often overlook the role of corpus complexity -- a critical factor, as recent work shows that both corpus size and document length strongly affect dense retrieval performance.
In this paper, we introduce CoRECT (\textbf{Co}ntrolled \textbf{R}etrieval \textbf{E}valuation of \textbf{C}ompression \textbf{T}echniques), a framework for large-scale evaluation of embedding compression methods, supported by a newly curated dataset collection.
To demonstrate its utility, we benchmark eight representative types of compression methods.
Notably, we show that non-learned compression achieves substantial index size reduction, even on up to 100M passages, with statistically insignificant performance loss.
However, selecting the optimal compression method remains challenging, as performance varies across models.
Such variability highlights the necessity of CoRECT to enable consistent comparison and informed selection of compression methods.
All code, data, and results are available on GitHub\footnote{\url{https://github.com/padas-lab-de/CoRECT}} and HuggingFace.\footnote{\url{https://huggingface.co/datasets/PaDaS-Lab/CoRE}}

\end{abstract}

\section{Introduction}

The success and wide-spread adoption of large language models (LLMs) has caused a shift in the field of information retrieval, moving from sparse, keyword-based matching to dense retrieval methods. Although effective, such systems require large amounts of memory to store embedding-based indices, which can easily surpass the size of the underlying data. To overcome memory-related scalability constraints, researchers have developed a range of embedding compression techniques that achieve significant size reductions with little loss in retrieval quality \cite{Jina-BinaryEmbeddings,Kusupati2022,Microsoft-IndexCompression,Mixedbread-BinaryMRL,MongoDB-BinaryQuantization,Weaviate-BinaryQuantization,Xiao2022,Yamada2021,Yoon2024,Zhan2021}. While these methods show strong results on individual corpora, their systematic evaluation across a broader range of models and datasets is still lacking. In addition, they overlook how retrieval performance scales with corpus complexity, which may cause an overestimation of effectiveness on large and heterogeneous collections~\cite{Luan2021,Reimers2021}, particularly when vector compression limits embedding expressiveness.

\begin{figure*}[t]
\centering
\includegraphics[width=\linewidth]{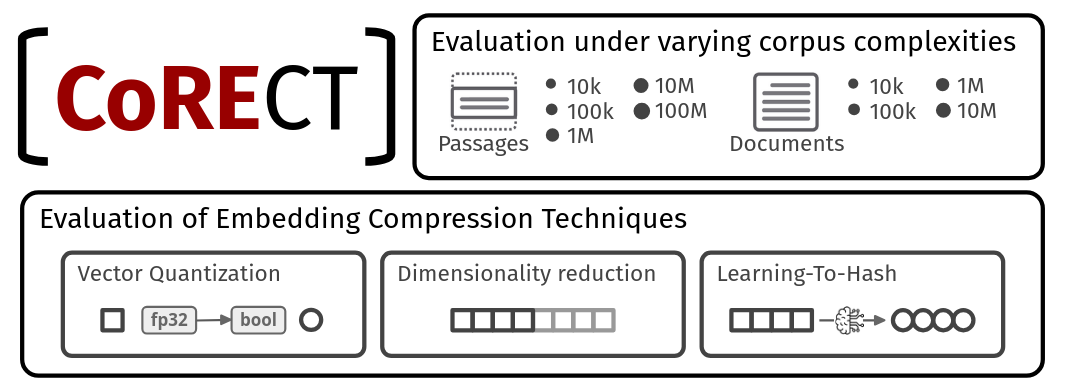}
\vspace{-0.8cm}
\caption{The CoRECT framework.}
\label{fig:corect_overview}
\vspace{-1em}
\end{figure*}

In this work, we critically re-examine recent claims regarding the effectiveness of embedding compression methods, including vector quantization, dimensionality reduction, and Learning-To-Hash (LTH) techniques. We introduce \textbf{CoRECT}, a framework for systematically evaluating and comparing compression methods across diverse datasets. Our main objective is to investigate how these techniques perform under varying levels of corpus complexity -- an aspect often overlooked in prior studies. To this end, we operationalize corpus complexity in two ways: (1) by scaling the corpus size from 10K to 100M passages and up to 10M documents, and (2) by comparing performance across both passage-level and document-level retrieval tasks (see Figure~\ref{fig:corect_overview}). To evaluate performance along these dimensions, we introduce a newly curated dataset collection named \textbf{Controlled Retrieval Evaluation (CoRE)}, constructed through targeted subsampling of MS MARCO v2\footnote{\url{https://microsoft.github.io/msmarco/TREC-Deep-Learning.html}} and comprising 76 human-judged TREC DL queries. CoRE serves as the primary evaluation dataset in CoRECT. To emphasize the robustness and generalizability of our findings, we further complement CoRE with datasets from BeIR \cite{Thakur2021-arxiv}. Finally, to demonstrate the utility of CoRECT, we apply the framework to a diverse and representative set of embedding compression methods. As such, our main contributions are as follows:

\begin{enumerate}
    \item The Controlled Retrieval Evaluation of Compression Techniques (CoRECT) framework, designed to systematically evaluate compression methods.
    To ensure robustness, we further extend our evaluation by including the domain-diverse corpora of the BeIR benchmark \cite{Thakur2021-arxiv}.
    \item The dataset collection CoRE, the cornerstone of CoRECT, is used to assess the effect of corpus complexity on retrieval performance.
    \item We apply CoRECT on four different models using eight representative types of compression techniques in over 40 combinations and find that choosing the wrong compression method can significantly impact retrieval performance. Furthermore, we show that there is no single best method that fits all models.
\end{enumerate}

\section{Related Work}

The following section reviews embedding compression methods, their evaluation, and related research on the influence of corpus complexity on performance.

\subsubsection{Compression Methods.}
\label{sec:lth}
The compression techniques discussed below fall into three broad categories. The first group reduces the precision of embeddings by representing each dimension with a lower number of bits. In the following, we will refer to this group of compression methods as Scalar and Binary Quantization (SBQ). It encompasses type-casting floating points, i.e. to float8~\cite{Micikevicius2022}, or mapping them to discrete integer values, i.e. \texttt{uint8}. Its simplicity makes SBQ a popular compression approach supported by various libraries and vector databases, such as FAISS \cite{Douze2024}, OpenSearch\footnote{\url{https://opensearch.org/}} or Milvus\footnote{\url{https://milvus.io/}}, as well as recent embedding models through quantization-aware training \cite{Voyage2025,Günther2025}.

The second group encompasses approaches for dimensionality reduction which remove certain dimensions from the embedding vector. The most common method in the context of data analytics is Principal Component Analysis (PCA), which can also be used to compress embedding vectors. More recently, Matryoshka Representation Learning (MRL)~\cite{Kusupati2022,Yoon2024}, which is integrated into model training, has found wide-spread adoption in contemporary embedding models~\cite{Voyage2025,Günther2025,Lee2025,Yu2024,Zhang2025}.

The third and last group consists of regular hashing and Learning-to-Hash (LTH) methods. A simple baseline is Locality-Sensitive Hashing (LSH), whose numerous proposed variants and extensions make it a well-studied compression methods~\cite{Bawa2005,Dasgupta2011,Datar2011,Gan2012,Huang2015}. For LTH methods, Product Quantization (PQ) \cite{Jégou2011} has emerged as a popular method with its own proposed expansions like Joint optimization of query encoding and Product Quantization (JPQ)~\cite{Zhan2021} or Distill-VQ \cite{Xiao2022}. Yamada et al. \cite{Yamada2021} introduce the Binary Passage Retriever (BPR) to binarize embeddings generated by the Dense Passage Retriever~\cite{Karpukhin2020}. Thakur et al.~\cite{Thakur2022} enhance prior work by introducing domain adaptation modules, enabling methods like BPR and JPQ to perform effectively in zero-shot retrieval settings.

\subsubsection{Evaluation of Compression Methods.}
Due to their popularity, SBQ methods have received growing attention in both academic studies and industry reports~\cite{Jina-BinaryEmbeddings,Microsoft-IndexCompression,Mixedbread-BinaryMRL,MongoDB-BinaryQuantization,Weaviate-BinaryQuantization}, demonstrating that high compression ratios can be achieved at minimal performance loss. Papers proposing new compression techniques tend to emphasize improvements over previous approaches~\cite{Thakur2022,Xiao2022,Yamada2021,Zhan2021}, comparing the retrieval performance of these compression techniques. However, most of these evaluations are limited to a small number of datasets and methods and do not consider scalability. To the best of our knowledge, the only thorough study comparing compression techniques was conducted by Zhang et al.~\cite{Zhang2023}, who evaluate them in the context of deep learning recommendation models. Additionally, they compare a small set of methods in a Retrieval Augmented Generation (RAG) setting on a single dataset, using exact match as a metric. However, due to their focus on recommendation, the chosen evaluation metrics and models are not representative of those commonly used in retrieval.

\subsubsection{Effect of Corpus Complexity.}
Recent work suggests that corpus complexity can have a significant impact on downstream effectiveness~\cite{Agrawal2023,Chen2024,Reimers2021}. Agrawal et al.~\cite{Agrawal2023} demonstrate that pretraining a language model on diverse and complex corpora improves their performance on downstream tasks. Reimers and Gurevych~\cite{Reimers2021} show that dense retrieval models struggle as corpus size increases. However, the only compression method the authors consider is vector truncation. Finally, Chen et al.~\cite{Chen2024} examine the effect of retrieval granularity, showing that the choice of retrieval unit, i.e. passage- or document-level, has a significant impact on performance, albeit without considering embedding compression. In this paper, we focus on exactly these two aspects of corpus complexity: a) the corpus size, which we increase from 10k to 100M and b) retrieval granularity, where we compare passage and document-level retrieval.

\section{Methodology}

In the following sections, we introduce the CoRECT framework, including the CoRE dataset collection, and demonstrate its utility for conducting a comprehensive and comparative analysis of embedding compression methods.
Table \ref{tab:models} lists the tested embedding models, including the number of parameters, dimensions and average performance on MMTEB \cite{Envoldsen2025}.
While Jina V3 is MRL trained on six different cutoff levels (32, 64, 128, 256, 512 and 768), Snowflake V2 is trained only at cutoff 256.

\begin{table}[t]
    \centering
    \caption{List of tested Embedding Models; Metric on MMTEB: NDCG@10 in \%}
    \label{tab:models}
    \begin{tabular}{@{}ccccc@{}}
        \toprule
        Model Name & \#Params & \#Dims & MRL & MMTEB \\ \midrule
        Jina V3~\cite{Sturua2024} & 572M & 1024 & \cmark & 55.76 \\
        Multilingual E5 (Large Instr.)~\cite{Wang2024} & 560M & 1024 & \xmark & 57.12 \\
        Snowflake-M V2~\cite{Yu2024} & 305M & 768 & \cmark & 54.83 \\
        Snowflake-M V1~\cite{Merrick2024} & 109M & 768 & \xmark & 39.33 \\
        \bottomrule
    \end{tabular}
\end{table}

\subsection{Dataset Creation and Subsampling} \label{dataset_creation}

The Controlled Retrieval Evaluation (CoRE) benchmark comprises two curated dataset collections -- passages and documents -- each derived from MS MARCO v2. Both collections are designed to enable controlled variation in the number of non-relevant corpus items included in the retrieval task. Hence, CoRE varies along two key dimensions -- corpus size and retrieval granularity -- while keeping the underlying retrieval task and relevance judgments fixed. The passage collection (average length: 286$\,\pm\,$111 characters) and the document collection (average length: 9\,010$\,\pm\,$15\,216 characters) are annotated with high-quality, human-judged relevance labels from the TREC Deep Learning 2023 campaign. For passage retrieval, CoRE includes 65 queries across five corpus sizes (10K, 100K, 1M, 10M, and 100M). For document retrieval, it offers 55 queries across four corpus sizes (10K to 10M). Each query is associated with exactly 10 relevant passages or documents, ensuring comparability across corpus scales.

A central challenge in constructing smaller subsets of large corpora is preserving a high density of meaningful distractors, as naive random sampling often removes these and renders the retrieval task unrealistically easy.
To address this issue, we adopt the intelligent subsampling strategy proposed by Fröbe et al.~\cite{Fröbe2025}, mining distractor documents from pooled ranking lists submitted to the 2023 TREC Deep Learning track.
To ensure quality, we discard the bottom 20\% of runs based on effectiveness and apply Reciprocal Rank Fusion~(RRF)~\cite{Cormack2009} to merge the top-ranked documents from the remaining runs.
For each query, the top 100 irrelevant or unjudged documents from this fused list are selected as high-quality distractors.
To simulate a realistic and challenging evaluation scenario, each query is paired with exactly 100 mined distractors, besides the 10 relevant documents.
The remainder of each corpus is filled with truly random documents that are neither relevant nor distractors under this definition.

\subsection{Framework}

The current implementation of the CoRECT framework evaluates compression methods on CoRE as well as the public BeIR datasets, tested with the four embedding models described above.
The included compression techniques encompass eight distinct types, which will be discussed in detail in the next subsection.

Given a model and dataset, embeddings are generated in batches and stored for downstream analysis.
Compression is then applied sequentially on each batch before performing retrieval using cosine similarity.
Any thresholds or parameters specific to the compression method are learned per batch, with the same values applied to both query and document embeddings.
For each query, CoRECT computes standard retrieval metrics such as NDCG, Recall, and MRR across multiple cutoff values $k$, and stores the results in JSON format.

The framework builds on widely used Python libraries, including Transformers~\cite{Wolf2020} for model instantiation and the Hugging Face Hub for data loading.
This modular design facilitates the integration of additional datasets, embedding models, and compression techniques.
New models available through Transformers can be incorporated by extending the base interface class for model loading and embedding compression, while new retrieval datasets from Hugging Face only require defining a custom loading function.
Comprehensive instructions for extending the framework are provided in the project’s README\footnote{\url{https://github.com/padas-lab-de/CoRECT}}.

\subsection{Compression Methods} \label{compression_methods}

In the following, we discuss the set of representative methods that have been initially included in the framework in more detail.

\subsubsection{Floating Point Casting.}
A straightforward approach at embedding compression is downcasting embeddings from their native precision to a lower-precision floating-point format.
We choose the native precision of each embedding model according to its reported Tensortype on Hugging Face.
Using PyTorch~\cite{Paszke2019}, we cast embedding tensors to FP16, BF16, and FP8.
As PyTorch provides multiple FP8 variants differing in mantissa and exponent length,\footnote{\url{https://dev-discuss.pytorch.org/t/float8-in-pytorch-1-x/1815}} we employ two FP8 types with three- and two-bit mantissas, respectively.

\subsubsection{Scalar and Binary Quantization (SBQ).}
To achieve higher compression ratios, we quantize embedding vectors to 8-, 4-, and 2-bit unsigned integer ranges using two quantization strategies.
The first, Equal Distance Binning, computes the per-dimension minimum and maximum of the embeddings, divides this range into $2^x$ equidistant bins, and maps each floating-point value to its corresponding bin index.
To reduce sensitivity to outliers, we clip values at the 2.5th and 97.5th percentiles before binning.
The second method, Percentile Binning, determines bin boundaries based on percentiles so that each bin contains approximately the same number of values, thereby ensuring a more balanced quantization.
For binary quantization, embeddings are mapped to 1-bit values using two thresholding schemes: zero and the per-dimension median.

\subsubsection{Dimensionality Reduction.}
We employ two approaches to reduce the dimensionality of embedding vectors: truncation along Matryoshka Representation Learning (MRL) cutoff points and Principal Component Analysis (PCA).
For truncation, only the first $x$ dimensions of each embedding are retained, with $x$ determined by the MRL training configurations of Jina V3 and Snowflake V2, enabling direct comparison with non-MRL-trained models such as E5 and Snowflake V1.
Since the Snowflake models have a smaller embedding size than Jina V3 and E5 (768 vs. 1024), two distinct sets of cutoff points are defined to preserve comparable dimensionality ratios.
MRL truncation is further combined with the floating-point casting and SBQ methods described above to obtain higher compression ratios.
For PCA, the implementation from the FAISS library~\cite{Douze2024} is used.
The sole hyperparameter -- the output dimension of the compressed vector -- is aligned with the cutoff values applied in MRL truncation.

\subsubsection{Hashing and Learning-to-Hash.}
The final group of methods includes Lo\-ca\-lity-Sensitive Hashing (LSH) and Product Quantization (PQ), both implemented using FAISS~\cite{Douze2024}.
LSH compresses embedding vectors into binary codes, controlled by a hyperparameter that defines the bit length of the resulting code.
Four configurations are evaluated, corresponding to compression ratios of $4\times$, $8\times$, $16\times$, and $32\times$.
PQ, as implemented in FAISS, applies k-means clustering with L2 distance to partition each embedding into subvectors, encoding each subspace with the chosen number of bits.
Six combinations of subvector count and code size are compared.
To maintain consistent compression ratios across models, two distinct hyperparameter sets are defined -- one for 768-dimensional and one for 1024-dimensional embeddings.

\section{Results}

\begin{figure*}[t]
\centering
\includegraphics[width=\linewidth]{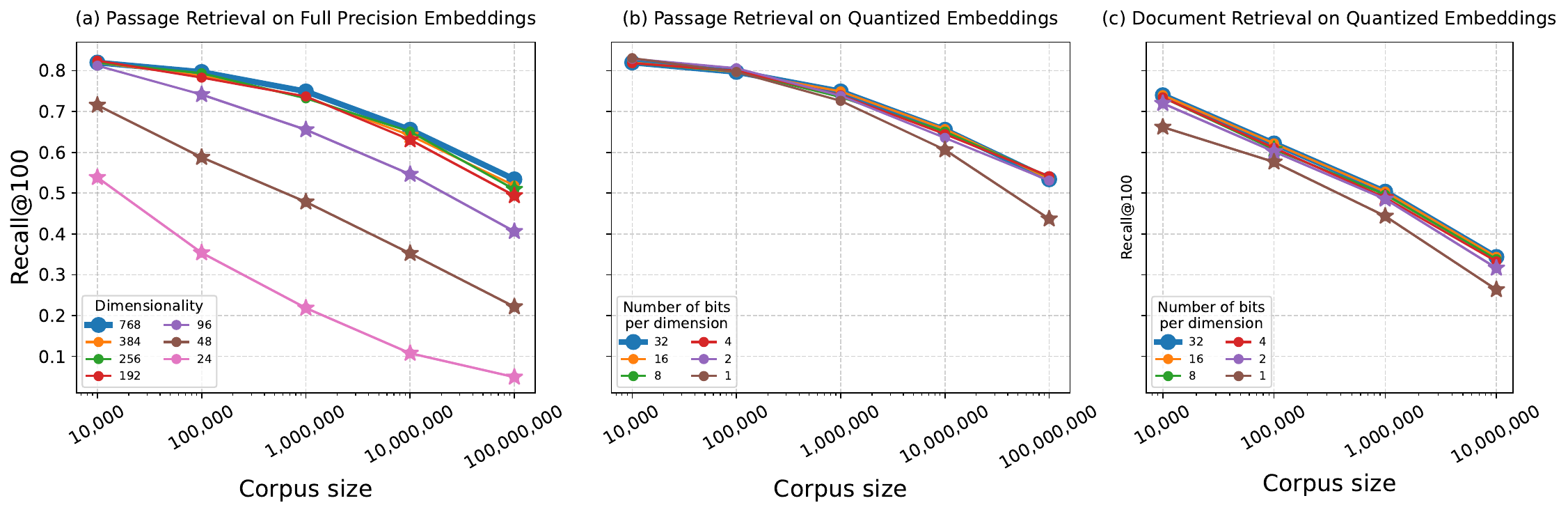}
\vspace{-0.6cm}
\caption{Recall@100 for Snowflake V2 across increasing corpus sizes in different compression settings. Stars indicate significantly lower recall than the blue baseline (one-sided Wilcoxon test, p=0.05). Recall decreases with higher corpus size and granularity. Quantization (middle) outperforms vector truncation (left).}
\label{fig:line_charts}
\vspace{-1em}
\end{figure*}

This section demonstrates the utility of the CoRECT framework by showcasing the analyses it enables.
Using CoRECT, we systematically compare the embedding compression methods introduced in Section~\ref{compression_methods} and examine how their performance varies across models, datasets, corpus sizes, and retrieval granularities.
These experiments highlight the framework’s ability to support controlled, large-scale evaluation of compression techniques -- insights that would be difficult to obtain without it.
Comprehensive results for all models, datasets, and methods are available on our project page.\footnote{\url{https://padas-lab-de.github.io/CoRECT/}}

\subsection{Impact of Corpus Complexity}

\textbf{Corpus Complexity Significantly Affects Retrieval Performance.} An important asset of our framework lies in its ability to evaluate the robustness of compression methods with respect to increasing corpus complexity. Figure~\ref{fig:line_charts} shows Recall@100 for Snowflake V2 on CoRE, measuring the model's ability to retrieve relevant documents among the top 100 results as a first-stage retriever. It consists of three line charts: (a) and (b) for passage retrieval, and (c) for document retrieval. In Plot (a), embedding vectors are only truncated at increasingly lower dimensions, while in Plot (b), only Percentile Binning is applied. Plot (c) mirrors the structure of (b) for document retrieval. The blue line, representing the uncompressed baseline, serves as a reference in the three charts. Across all settings, the retrieval performance decreases significantly as the corpus grows, even for the uncompressed baseline. This trend reflects the increasing difficulty of the retrieval task with larger passage or document collections.

\textbf{Scalability of Vector Truncation Depends on Compression Ratio.} Plot (a) isolates the impact of vector truncation on performance, reaching a maximum compression ratio of 32 when only the first 24 dimensions of each vector are retained. While moderate truncation to 384 or 256 dimensions barely affects the recall scores, significant performance differences, albeit small, start to manifest at 192 dimensions ($4\times$ compression) for the larger corpora. At a compression ratio of 8, i.e. 96 dimensions, recall starts to drop sharply, even for the 100k corpus. While, here, we only present the results for Snowflake V2, as it is one of the MRL trained models, we observed even better results for Jina V3, as the vector can be truncated to 256 dimensions ($4\times$ compression) without significant performance differences. Furthermore, truncating to a compression ratio of 8 only leads to a slight decrease in performance. Naturally, for Snowflake V1 and E5, the two models without any MRL training, truncation performs worse, quickly leading to significant degradations of recall as corpus size increases. Notably, for both models taking only half of the vector dimensions already has a significant performance impact when the corpus size increases to 1M.

\setlength{\tabcolsep}{4pt}
\begin{table*}[t]
\caption{Recall@100 (in \%) with increasing corpus size. The best result per column is marked in bold, the second best is underlined. The degree of performance degradation when increasing corpus complexity varies between models, making some more suitable for application on large corpora than others.}
\label{tab:performance}
\centering
\begin{tabular}{@{}lccccccccccccc@{}}
\toprule
 & \multicolumn{3}{c}{Passages} &  & \multicolumn{3}{c}{Documents} \\ \midrule
 & 10k & 1M & 100M &  & 10k & 1M & 10M \\ \midrule
Jina V3 & \underline{83.85} & 71.39 & 44.15 &  & 72.91 & 47.64 & 26.91 \\
E5 & \textbf{84.15} & \textbf{75.39} & 50.77 &  & \textbf{76.00} & 47.64 & 31.64 \\
Snowflake V2 & 82.00 & \underline{74.92} & \textbf{53.39} &  & \underline{74.00} & \underline{50.18} & \underline{34.36} \\
Snowflake V1 & 80.31 & 71.69 & \underline{50.92} &  & \underline{74.00} & \textbf{50.36} & \textbf{36.73} \\
\bottomrule
\end{tabular}
\end{table*}

\textbf{Quantization Outperforms Vector Truncation.} Plot (b) reveals that scalar quantization barely impacts performance, while binarizing the vector has a significant impact on recall performance at larger corpus sizes. Comparing the performance of quantization to that of vector truncation in plot (a) at the same compression ratio, we find that quantization has a clear edge. Looking at the remaining models, whose results are not shown in the plot, we observe an even clearer advantage for E5. For Jina V3, the performance difference between quantization and vector truncation becomes smaller due to the model's ability to retain high recall even with large vector truncation. Nevertheless, for both E5 and Jina V3 the quantized vectors retain most of the originals' performance. In contrast, quantizing Snowflake V1 embeddings causes a large degradation in recall, even though SBQ still performs slightly better than truncation. Considering retrieval granularity, plot (c) shows similar trends to plot (b) for document retrieval, albeit with faster overall performance degradation. This pattern can also be observed for the remaining models, where quantization performance trends on passage level translate to document retrieval.

\begin{table*}[t]
\caption{Recall@100 (in \%) for different corpus complexity with full and binary precision (no truncation). Choosing the wrong thresholding type can greatly decrease performance.}
\label{tab:binary}
\centering
\begin{tabular}{@{}lccccccccccccc@{}}
\toprule
 & \multicolumn{2}{c}{Jina V3} &  & \multicolumn{2}{c}{E5} &  & \multicolumn{2}{c}{Snowflake V2} &  & \multicolumn{2}{c}{Snowflake V1} \\ \midrule
 & 10k & 100M &  & 10k & 100M &  & 10k & 100M &  & 10k & 100M \\ \midrule
Full precision & 83.85 & 44.15 &  & 84.31 & 50.92 &  & 82.00 & 53.39 &  & 80.31 & 50.92 \\ \midrule
Zero Thresh. & \textbf{83.39} & \textbf{41.39} &  & 71.23 & 19.23 &  & \textbf{82.62} & \textbf{49.39} &  & \textbf{78.92} & \textbf{32.46} \\
Median Thresh. & 82.92 & 35.69 &  & \textbf{86.31} & \textbf{39.54} &  & \textbf{82.62} & 43.39 &  & 70.77 & 18.00 \\
\bottomrule
\end{tabular}
\end{table*}

\textbf{Model Robustness to Corpus Complexity Varies.} Apart from allowing us to evaluate the robustness of compression methods with respect to corpus size and retrieval granularity, CoRE also enables us to study a model's general capability of dealing with these challenges. Interestingly, we can observe different degrees of performance degradation depending on the model. As shown in Table \ref{tab:performance}, while E5 performs best on the small 10k passage corpus followed by Jina V3, on the largest 100M corpus Snowflake V2 takes the lead with E5 and Jina V3 in the last two places. For document retrieval, E5 again takes the lead for 10K, while for the larger corpora Snowflake V1, the smallest model, demonstrates the best recall and Jina V3, the largest one, the worst. As the embedding models demonstrate different capability of retaining recall with increasing corpus complexity, we demonstrate the general importance of evaluating model performance on large-scale datasets.

\subsection{Compression Robustness Across Datasets and Models}

\textbf{Choosing the Wrong Compression Method Harms Performance.}
When comparing the performance of methods with an equal compression ratio across models, we find that specific model-compression method combinations significantly reduce retrieval performance. Table \ref{tab:binary} presents Recall@100 for each model, comparing full-precision baselines with two binary quantization methods: zero and median thresholding. Interestingly, zero thresholding outperforms median thresholding for three of the four models, particularly on the 100M corpus. The larger discrepancy at 100M is most likely a result of the median being estimated on a limited sample rather than the full dataset. Despite its simplicity, zero thresholding achieves remarkably strong performance, with E5 being the only model where this approach leads to a notable drop in recall. In this paper, we restrict ourselves to reporting these empirical observations; a deeper interpretation of the underlying causes is left for future work.

\begin{figure}[t!]
    \centering
    \begin{subfigure}{0.49\linewidth}
        \centering
        \includegraphics[width=\linewidth]{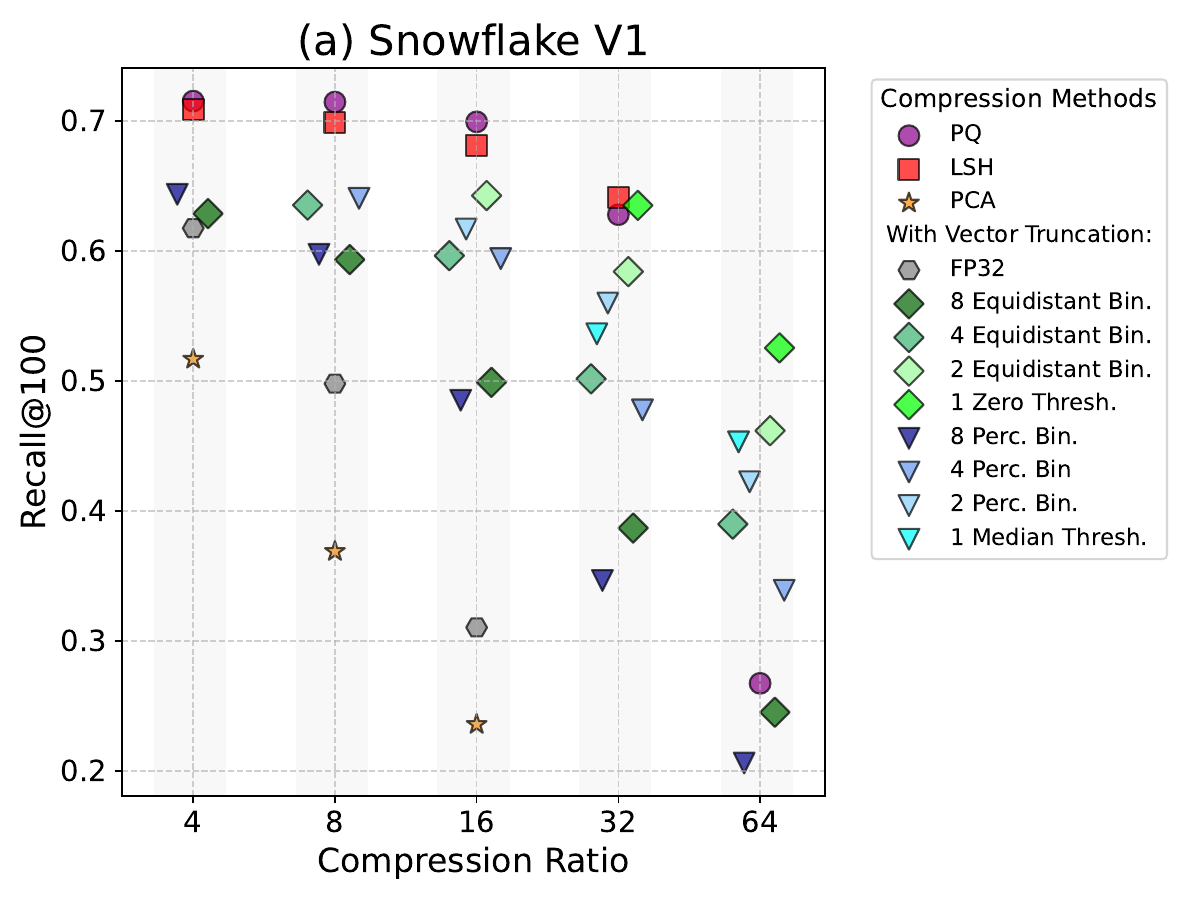}
    \end{subfigure}
    \hfill
    \begin{subfigure}{0.49\linewidth}
        \centering
        \includegraphics[width=\linewidth]{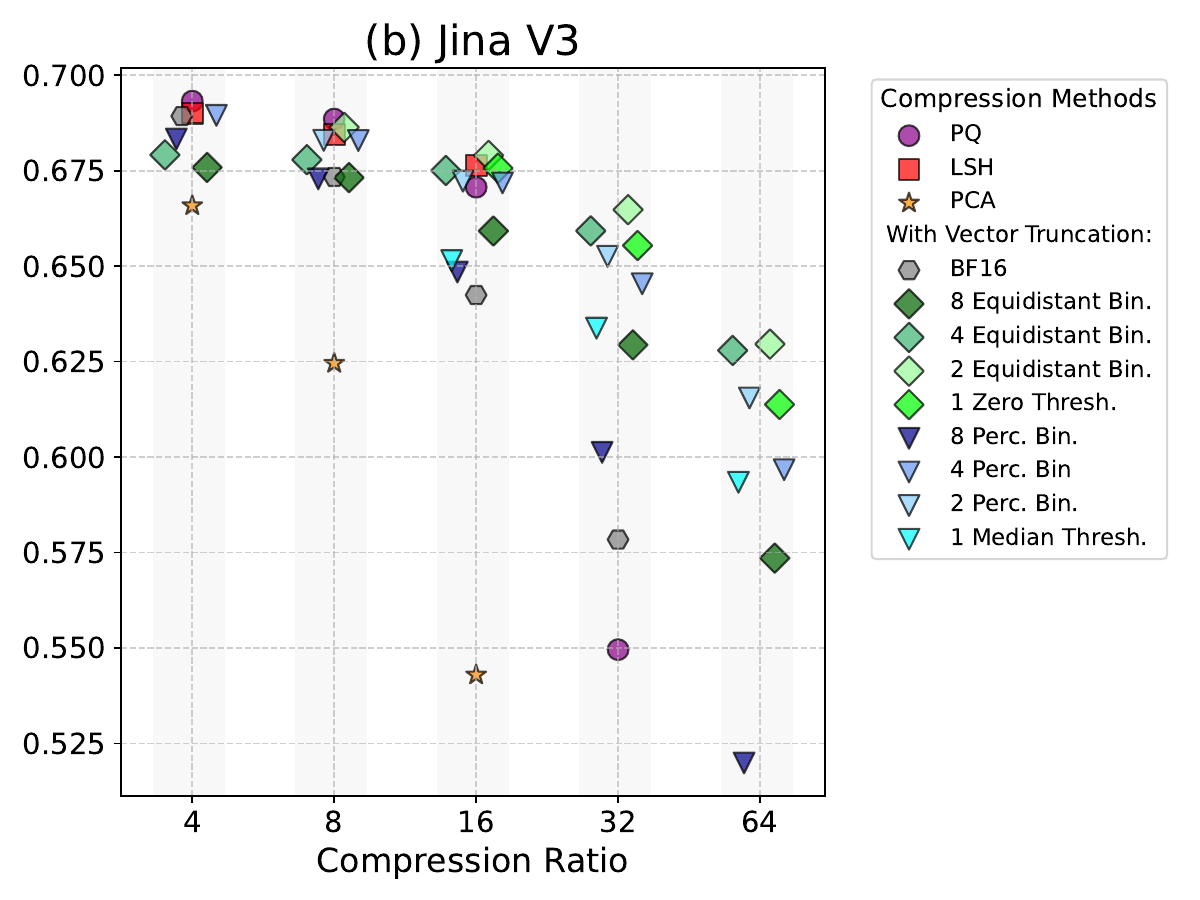}
    \end{subfigure}
    \vspace{-0.2cm}
    \caption{Average Recall@100 across BeIR datasets for different compression methods. The x-values of some compression methods have been shifted slightly for better visibility. The light gray columns in the background mark the compression ratio to which the methods belong. Shapes indicate different types of compression techniques. Some methods have been combined with vector truncation to achieve the desired compression ratio as indicated in the legend. The top-performing measures vary depending on the model.}
    \label{fig:pareto_plots}
    \vspace{-1em}
\end{figure}

\textbf{Large Compression Ratios Show Differences of Methods.}
Expanding our focus to evaluating the robustness of compression methods on a diverse set of corpora, we apply CoRECT on the BeIR datasets.
Although the ranking of compression methods changes between datasets, these variations are largely attributable to minor performance differences among densely clustered methods.
Figure~\ref{fig:pareto_plots} presents average Recall@100 across BeIR for different compression ratios.
For Jina V3 (Plot b), the methods cluster closely at low compression rates, but clear performance differences emerge as the compression ratio increases.
Quantization to 2 bits performs robustly across all ratios—a trend also observed for Snowflake V2 and E5.
Within the SBQ family, combining moderate vector truncation with quantization yields high compression ratios while often improving retrieval performance compared to heavier quantization of the full vector.
This approach works particularly well with Jina’s MRL-training and, to a lesser extent, for Snowflake V2 and E5 at smaller truncation levels.
An exception is the Snowflake V1 model, whose performance degrades sharply as compression increases (see Plot a).
Although LSH and PQ perform relatively well at lower ratios, their effectiveness drops substantially at $32\times$ compression.
Across most settings, however, PQ and LSH still outperform scalar quantization, making them a better choice for compressing Snowflake V1 embeddings than SBQ.

\begin{wrapfigure}{r}{0.5\textwidth}
    \vspace{-1.5em}
    \includegraphics[width=\linewidth]{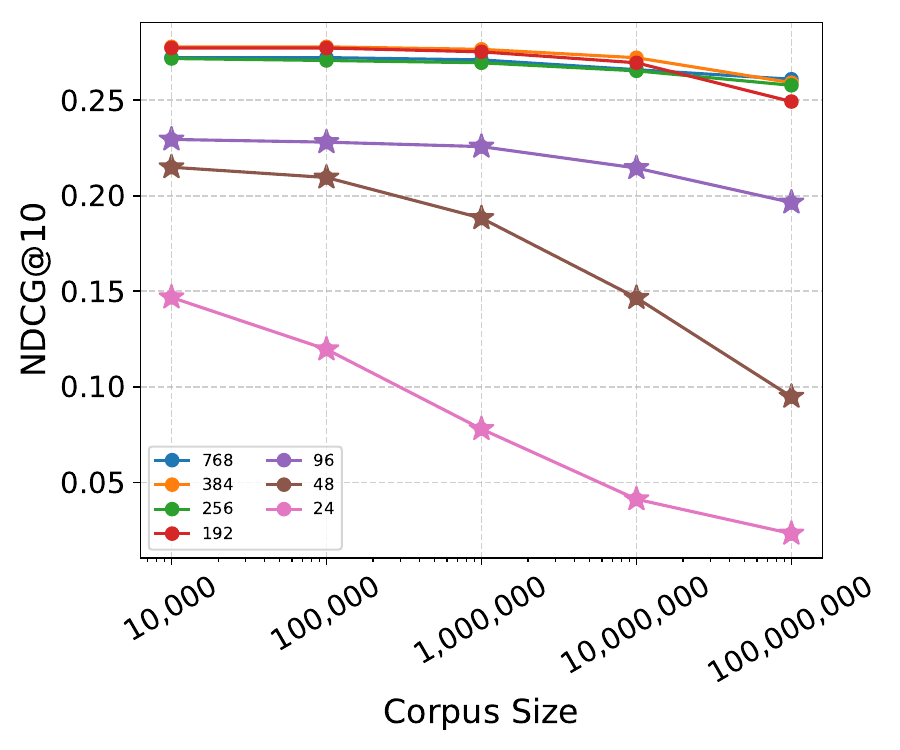}
    \vspace{-0.8cm}
    \caption{NDCG@10 for Snowflake V2 on the CoRE passage corpora. In contrast to Recall@100, NDCG@10 remains stable with increasing corpus size.}
    \label{fig:ndcg}
    \vspace{-5em}
\end{wrapfigure}

\begin{wrapfigure}{r}{0.5\textwidth}
    \vspace{-1.5em}
    \includegraphics[width=\linewidth,trim={10cm 0cm 10cm 0cm},clip]{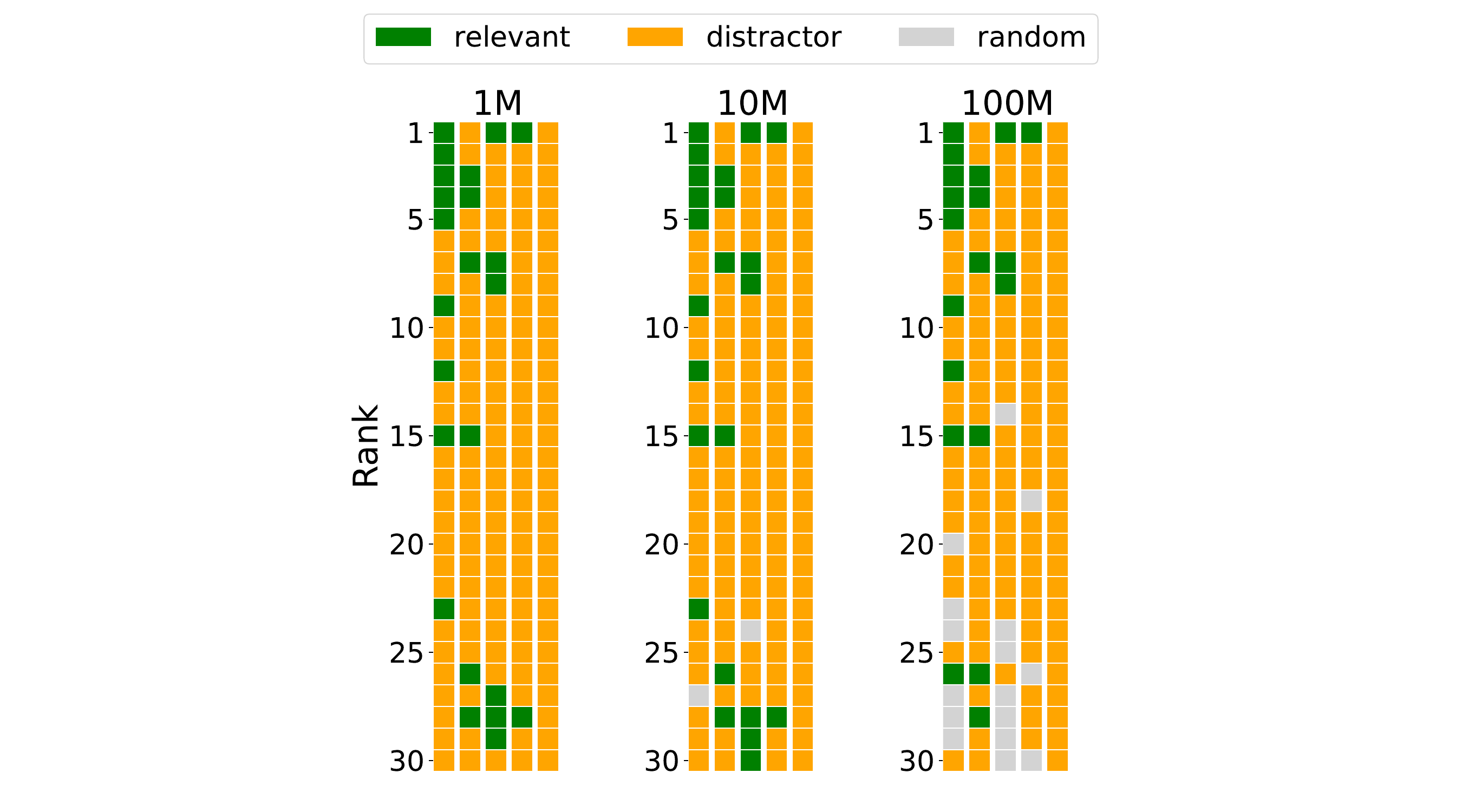}
    \vspace{-0.8cm}
    \caption{Positions of retrieved document types for Snowflake V2 on three CoRE passage corpora for five queries. While the top-10 ranking remains stable, random documents start to appear at lower ranks as corpus size increases, pushing relevant documents down in the ranking.}
    \label{fig:rankings}
    \vspace{-5em}
\end{wrapfigure}

\textbf{Type of Retrieval Task Affects Compression Performance}.
So far, we only considered compression in the context of first-stage retrieval, using Recall@100 as a performance metric.
However, when using an embedding model as a single-stage retrieval system, NDCG@10 becomes a more suitable metric, as the number of retrieved documents will be lower.
Figure \ref{fig:ndcg} shows that in contrast to the obvious performance degradation of Recall@100 in Figure \ref{fig:line_charts}, NDCG@10 remains far more stable with increasing corpus size.
This effect reflects the higher difficulty for a random document embedding to achieve a level of similarity that is high enough to retrieve it in a top-10 position.
As an example, Figure \ref{fig:rankings} shows the ranks at which different types of documents (relevant, distractor or random) are retrieved as corpus size increases from 1M to 100M when truncating to 192 dimensions.
Each column represents retrieval results for one specific query.
While the top-30 ranks for the 1M corpus do not contain a single random document, they start appearing for the 10M corpus and manage to push some relevant documents out of the top-30 in the 100M corpus.
Notably, the top-10 results remain stable as corpus size increases.
As such, when considering only the top-10 results, stronger compression is possible without significantly reducing retrieval performance.
This demonstrates that the primary metric for evaluating compression methods should be chosen according to the retrieval context.

Overall, our results demonstrate that there is no single best compression method that fits all models. Rather, the optimal choice is model-dependent, necessitating a thorough evaluation and comparison of compression methods. Thus, comparing compression techniques across diverse corpora allows us to a) identify methods that consistently perform well and b) exclude methods that are unsuitable for a particular model.

\section{Discussion and Future Work}

Overall, the insights generated with the CoRECT framework enable us to derive certain guidelines for applying compression techniques in practice:

\begin{enumerate}
    \item \textbf{Evaluate model performance on large datasets.} When choosing a model to generate millions of embeddings, it is essential to evaluate its performance on large corpora. As our analysis shows, some models' retrieval performance degrades quicker when increasing corpus size, which can lead to a significant decrease in performance when choosing the wrong model.
    \item \textbf{Compare Compression Performance of Different Methods.} When choosing a compression method, comparing its performance to a diverse set of other methods at a fixed compression ratio is important, as our analysis has demonstrated that a method that performs well for one model does not necessarily do so for another. Selecting a method that is unsuitable for a specific model can greatly impact performance.
    \item \textbf{Try Combining Compression Techniques.} For three out of our four models, combining quantization with moderate vector truncation achieved the best retrieval performance at a fixed compression ratio. While the number of method combinations we considered is fairly limited, the results still highlight the potential benefit of mixing different methods.
\end{enumerate}

While our framework aims to compare a wide variety of compression methods, a clear limitation lies in its current focus on SBQ along with techniques for dimensionality reduction, ignoring more recent LTH approaches described in Section \ref{sec:lth}. To demonstrate the initial capabilities of our framework, we focused on popular compression techniques that are not as train-intensive as recent LTH methods. However, we acknowledge that integrating such methods in the future would be a valuable extension of CoRECT, allowing the application and evaluation of these methods on modern embedding models. Furthermore, we only evaluate compression techniques based on their performance, ignoring other factors like their storage efficiency, computation time and retrieval speed.

\section{Conclusion}

We introduced CoRECT, a framework designed to robustly evaluate the performance of embedding compression methods with respect to (1) corpus complexity and (2) consistency across diverse corpora.
To study the first aspect, we developed CoRE, a benchmark that enables controlled variation in the number of non-relevant passages or documents included in the retrieval task.
To address the second aspect, we leverage the diverse BeIR datasets.

Applying CoRECT to a representative set of compression techniques -- scalar and binary quantization, dimensionality reduction, locality-sensitive hashing, and product quantization, among others -- across four different retrieval models reveals that there is no “one-size-fits-all” compression method that performs uniformly well across models and datasets.
Crucially, these findings emerge only through CoRECT’s systematic and controlled evaluation design, which is explicitly tailored to disentangle the effects of corpus complexity and model-specific behavior.
Through this framework, we uncover model-dependent pitfalls, where certain compression techniques significantly degrade performance for one model while remaining effective for others.

Using CoRE, we further show that retrieval performance generally declines with increasing corpus complexity.
While this trend is expected, the degree of degradation varies substantially across models, suggesting that some are inherently more robust to scaling than others.
Overall, CoRECT demonstrates the necessity of model-specific, controlled evaluation when applying embedding compression techniques, enabling a more nuanced and reliable understanding of their real-world retrieval performance.

\begin{credits}
\section*{Acknowledgments}
\begin{minipage}{0.30\linewidth}
\centering
\includegraphics[width=0.9\textwidth]{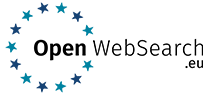}
\end{minipage}
\begin{minipage}{0.67\linewidth}
\href{https://doi.org/10.3030/101070014}{This work has received funding from the European Union's Horizon Europe research and innovation program under grant agreement No. 101070014 (OpenWebSearch.EU)}.
\end{minipage}

\vspace{0.3cm}
\noindent Furthermore, this work was funded by the Bavarian State Ministry of Economic Affairs, Regional Development, and Energy (StMWi) and is supported by funds of the Federal Ministry of Agriculture, Food and Regional Identity (BMLEH) based on a decision of the Parliament of the Federal Republic of Germany via the Federal Office for Agriculture and Food (BLE) under the strategy for digitalisation in agriculture.
\end{credits}

\bibliographystyle{splncs04}
\bibliography{sources}

@inproceedings{Chen2024,
    title = "Dense {X} Retrieval: What Retrieval Granularity Should We Use?",
    author = "Chen, Tong  and Wang, Hongwei  and Chen, Sihao  and Yu, Wenhao  and Ma, Kaixin  and Zhao, Xinran  and Zhang, Hongming  and Yu, Dong",
    editor = "Al-Onaizan, Yaser  and Bansal, Mohit  and Chen, Yun-Nung",
    booktitle = "Proceedings of the 2024 Conference on Empirical Methods in Natural Language Processing",
    month = nov,
    year = "2024",
    address = "Miami, Florida, USA",
    publisher = "Association for Computational Linguistics",
    url = "https://aclanthology.org/2024.emnlp-main.845/",
    doi = "10.18653/v1/2024.emnlp-main.845",
    pages = "15159--15177",
}

@inproceedings{Agrawal2023,
  title={Corpus complexity matters in pretraining language models},
  author={Agrawal, Ameeta and Singh, Suresh},
  booktitle={Proceedings of The Fourth Workshop on Simple and Efficient Natural Language Processing (SustaiNLP)},
  pages={257--263},
  year={2023}
}

@article{Zhang2023,
    author = {Zhang, Hailin and Zhao, Penghao and Miao, Xupeng and Shao, Yingxia and Liu, Zirui and Yang, Tong and Cui, Bin},
    title = {Experimental Analysis of Large-Scale Learnable Vector Storage Compression},
    year = {2023},
    issue_date = {December 2023},
    publisher = {VLDB Endowment},
    volume = {17},
    number = {4},
    issn = {2150-8097},
    url = {https://doi.org/10.14778/3636218.3636234},
    doi = {10.14778/3636218.3636234},
    journal = {Proc. VLDB Endow.},
    month = dec,
    pages = {808–822},
    numpages = {15}
}

@inproceedings{Wolf2020,
    title = "Transformers: State-of-the-Art Natural Language Processing",
    author = "Thomas Wolf and Lysandre Debut and Victor Sanh and Julien Chaumond and Clement Delangue and Anthony Moi and Pierric Cistac and Tim Rault and Rémi Louf and Morgan Funtowicz and Joe Davison and Sam Shleifer and Patrick von Platen and Clara Ma and Yacine Jernite and Julien Plu and Canwen Xu and Teven Le Scao and Sylvain Gugger and Mariama Drame and Quentin Lhoest and Alexander M. Rush",
    booktitle = "Proceedings of the 2020 Conference on Empirical Methods in Natural Language Processing: System Demonstrations",
    month = oct,
    year = "2020",
    address = "Online",
    publisher = "Association for Computational Linguistics",
    url = "https://www.aclweb.org/anthology/2020.emnlp-demos.6",
    pages = "38--45"
}

@inproceedings{Yamada2021,
    title = "Efficient Passage Retrieval with Hashing for Open-domain Question Answering",
    author = "Yamada, Ikuya  and
      Asai, Akari  and
      Hajishirzi, Hannaneh",
    editor = "Zong, Chengqing  and
      Xia, Fei  and
      Li, Wenjie  and
      Navigli, Roberto",
    booktitle = "Proceedings of the 59th Annual Meeting of the Association for Computational Linguistics and the 11th International Joint Conference on Natural Language Processing (Volume 2: Short Papers)",
    month = aug,
    year = "2021",
    address = "Online",
    publisher = "Association for Computational Linguistics",
    url = "https://aclanthology.org/2021.acl-short.123/",
    doi = "10.18653/v1/2021.acl-short.123",
    pages = "979--986"
}

@inproceedings{Zhan2021,
author = {Zhan, Jingtao and Mao, Jiaxin and Liu, Yiqun and Guo, Jiafeng and Zhang, Min and Ma, Shaoping},
title = {Jointly Optimizing Query Encoder and Product Quantization to Improve Retrieval Performance},
year = {2021},
isbn = {9781450384469},
publisher = {Association for Computing Machinery},
address = {New York, NY, USA},
url = {https://doi.org/10.1145/3459637.3482358},
doi = {10.1145/3459637.3482358},
pages = {2487–2496},
numpages = {10},
location = {Virtual Event, Queensland, Australia},
series = {CIKM '21}
}

@misc{Thakur2022,
  doi = {10.48550/ARXIV.2205.11498},
  url = {https://arxiv.org/abs/2205.11498},
  author = {Thakur, Nandan and Reimers, Nils and Lin, Jimmy},
  title = {Injecting Domain Adaptation with Learning-to-hash for Effective and Efficient Zero-shot Dense Retrieval},
  publisher = {arXiv},
  year = {2022},
  copyright = {Creative Commons Attribution Share Alike 4.0 International}
}

@inproceedings{Xiao2022,
author = {Xiao, Shitao and Liu, Zheng and Han, Weihao and Zhang, Jianjin and Lian, Defu and Gong, Yeyun and Chen, Qi and Yang, Fan and Sun, Hao and Shao, Yingxia and Xie, Xing},
title = {Distill-VQ: Learning Retrieval Oriented Vector Quantization By Distilling Knowledge from Dense Embeddings},
year = {2022},
isbn = {9781450387323},
publisher = {Association for Computing Machinery},
address = {New York, NY, USA},
url = {https://doi.org/10.1145/3477495.3531799},
doi = {10.1145/3477495.3531799},
booktitle = {Proceedings of the 45th International ACM SIGIR Conference on Research and Development in Information Retrieval},
pages = {1513–1523},
numpages = {11},
location = {Madrid, Spain},
series = {SIGIR '22}
}

@inproceedings{Yoon2024,
    title = "Matryoshka-Adaptor: Unsupervised and Supervised Tuning for Smaller Embedding Dimensions",
    author = "Yoon, Jinsung and
      Sinha, Rajarishi and
      Arik, Sercan O and
      Pfister, Tomas",
    editor = "Al-Onaizan, Yaser  and
      Bansal, Mohit  and
      Chen, Yun-Nung",
    booktitle = "Proceedings of the 2024 Conference on Empirical Methods in Natural Language Processing",
    month = nov,
    year = "2024",
    address = "Miami, Florida, USA",
    publisher = "Association for Computational Linguistics",
    url = "https://aclanthology.org/2024.emnlp-main.576/",
    doi = "10.18653/v1/2024.emnlp-main.576",
    pages = "10318--10336"
}

@inproceedings{Kusupati2022,
 author = {Kusupati, Aditya and Bhatt, Gantavya and Rege, Aniket and Wallingford, Matthew and Sinha, Aditya and Ramanujan, Vivek and Howard-Snyder, William and Chen, Kaifeng and Kakade, Sham and Jain, Prateek and Farhadi, Ali},
 booktitle = {Advances in Neural Information Processing Systems},
 editor = {S. Koyejo and S. Mohamed and A. Agarwal and D. Belgrave and K. Cho and A. Oh},
 pages = {30233--30249},
 publisher = {Curran Associates, Inc.},
 title = {Matryoshka Representation Learning},
 volume = {35},
 year = {2022}
}

@misc{Günther2025,
      title={jina-embeddings-v4: Universal Embeddings for Multimodal Multilingual Retrieval}, 
      author={Michael Günther and Saba Sturua and Mohammad Kalim Akram and Isabelle Mohr and Andrei Ungureanu and Sedigheh Eslami and Scott Martens and Bo Wang and Nan Wang and Han Xiao},
      year={2025},
      eprint={2506.18902},
      archivePrefix={arXiv},
      primaryClass={cs.AI},
      url={https://arxiv.org/abs/2506.18902}, 
}

@article{Voyage2025,
  author = {Voyage AI},
  title = {voyage-3-large: the new state-of-the-art general-purpose embedding model},
  journal = {Voyage AI Blog},
  year = {2025},
  note = {https://blog.voyageai.com/2025/01/07/voyage-3-large/},
}

@misc{Micikevicius2022,
      title={FP8 Formats for Deep Learning}, 
      author={Paulius Micikevicius and Dusan Stosic and Neil Burgess and Marius Cornea and Pradeep Dubey and Richard Grisenthwaite and Sangwon Ha and Alexander Heinecke and Patrick Judd and John Kamalu and Naveen Mellempudi and Stuart Oberman and Mohammad Shoeybi and Michael Siu and Hao Wu},
      year={2022},
      eprint={2209.05433},
      archivePrefix={arXiv},
      primaryClass={cs.LG},
      url={https://arxiv.org/abs/2209.05433}, 
}

@article{Douze2024,
      title={The Faiss library},
      author={Matthijs Douze and Alexandr Guzhva and Chengqi Deng and Jeff Johnson and Gergely Szilvasy and Pierre-Emmanuel Mazaré and Maria Lomeli and Lucas Hosseini and Hervé Jégou},
      year={2024},
      eprint={2401.08281},
      archivePrefix={arXiv},
      primaryClass={cs.LG}
}

@article{Zhang2025,
  title={Qwen3 Embedding: Advancing Text Embedding and Reranking Through Foundation Models},
  author={Zhang, Yanzhao and Li, Mingxin and Long, Dingkun and Zhang, Xin and Lin, Huan and Yang, Baosong and Xie, Pengjun and Yang, An and Liu, Dayiheng and Lin, Junyang and Huang, Fei and Zhou, Jingren},
  journal={arXiv preprint arXiv:2506.05176},
  year={2025}
}

@misc{Yu2024,
      title={Arctic-Embed 2.0: Multilingual Retrieval Without Compromise}, 
      author={Puxuan Yu and Luke Merrick and Gaurav Nuti and Daniel Campos},
      year={2024},
      eprint={2412.04506},
      archivePrefix={arXiv},
      primaryClass={cs.CL},
      url={https://arxiv.org/abs/2412.04506}, 
}

@misc{Lee2025,
      title={Gemini Embedding: Generalizable Embeddings from Gemini}, 
      author={Jinhyuk Lee and Feiyang Chen and Sahil Dua and Daniel Cer and Madhuri Shanbhogue and Iftekhar Naim and Gustavo Hernández Ábrego and Zhe Li and Kaifeng Chen and Henrique Schechter Vera and Xiaoqi Ren and Shanfeng Zhang and Daniel Salz and Michael Boratko and Jay Han and Blair Chen and Shuo Huang and Vikram Rao and Paul Suganthan and Feng Han and Andreas Doumanoglou and Nithi Gupta and Fedor Moiseev and Cathy Yip and Aashi Jain and Simon Baumgartner and Shahrokh Shahi and Frank Palma Gomez and Sandeep Mariserla and Min Choi and Parashar Shah and Sonam Goenka and Ke Chen and Ye Xia and Koert Chen and Sai Meher Karthik Duddu and Yichang Chen and Trevor Walker and Wenlei Zhou and Rakesh Ghiya and Zach Gleicher and Karan Gill and Zhe Dong and Mojtaba Seyedhosseini and Yunhsuan Sung and Raphael Hoffmann and Tom Duerig},
      year={2025},
      eprint={2503.07891},
      archivePrefix={arXiv},
      primaryClass={cs.CL},
      url={https://arxiv.org/abs/2503.07891}, 
}

@misc{Karpukhin2020,
      title={Dense Passage Retrieval for Open-Domain Question Answering}, 
      author={Vladimir Karpukhin and Barlas Oğuz and Sewon Min and Patrick Lewis and Ledell Wu and Sergey Edunov and Danqi Chen and Wen-tau Yih},
      year={2020},
      eprint={2004.04906},
      archivePrefix={arXiv},
      primaryClass={cs.CL},
      url={https://arxiv.org/abs/2004.04906}, 
}

@ARTICLE{Jégou2011,
  author={Jégou, Herve and Douze, Matthijs and Schmid, Cordelia},
  journal={IEEE Transactions on Pattern Analysis and Machine Intelligence}, 
  title={Product Quantization for Nearest Neighbor Search}, 
  year={2011},
  volume={33},
  number={1},
  pages={117-128},
  doi={10.1109/TPAMI.2010.57}
}

@inproceedings{Bawa2005,
author = {Bawa, Mayank and Condie, Tyson and Ganesan, Prasanna},
title = {LSH forest: self-tuning indexes for similarity search},
year = {2005},
isbn = {1595930469},
publisher = {Association for Computing Machinery},
address = {New York, NY, USA},
url = {https://doi.org/10.1145/1060745.1060840},
doi = {10.1145/1060745.1060840},
booktitle = {Proceedings of the 14th International Conference on World Wide Web},
pages = {651–660},
numpages = {10},
location = {Chiba, Japan},
series = {WWW '05}
}

@inproceedings{Gan2012,
author = {Gan, Junhao and Feng, Jianlin and Fang, Qiong and Ng, Wilfred},
title = {Locality-sensitive hashing scheme based on dynamic collision counting},
year = {2012},
isbn = {9781450312479},
publisher = {Association for Computing Machinery},
address = {New York, NY, USA},
url = {https://doi.org/10.1145/2213836.2213898},
doi = {10.1145/2213836.2213898},
booktitle = {Proceedings of the 2012 ACM SIGMOD International Conference on Management of Data},
pages = {541–552},
numpages = {12},
location = {Scottsdale, Arizona, USA},
series = {SIGMOD '12}
}

@inproceedings{Datar2011,
author = {Datar, Mayur and Immorlica, Nicole and Indyk, Piotr and Mirrokni, Vahab S.},
title = {Locality-sensitive hashing scheme based on p-stable distributions},
year = {2004},
isbn = {1581138857},
publisher = {Association for Computing Machinery},
address = {New York, NY, USA},
url = {https://doi.org/10.1145/997817.997857},
doi = {10.1145/997817.997857},
booktitle = {Proceedings of the Twentieth Annual Symposium on Computational Geometry},
pages = {253–262},
numpages = {10},
location = {Brooklyn, New York, USA},
series = {SCG '04}
}

@inproceedings{Dasgupta2011,
author = {Dasgupta, Anirban and Kumar, Ravi and Sarlos, Tamas},
title = {Fast locality-sensitive hashing},
year = {2011},
isbn = {9781450308137},
publisher = {Association for Computing Machinery},
address = {New York, NY, USA},
url = {https://doi.org/10.1145/2020408.2020578},
doi = {10.1145/2020408.2020578},
booktitle = {Proceedings of the 17th ACM SIGKDD International Conference on Knowledge Discovery and Data Mining},
pages = {1073–1081},
numpages = {9},
location = {San Diego, California, USA},
series = {KDD '11}
}

@article{Huang2015,
author = {Huang, Qiang and Feng, Jianlin and Zhang, Yikai and Fang, Qiong and Ng, Wilfred},
title = {Query-aware locality-sensitive hashing for approximate nearest neighbor search},
year = {2015},
issue_date = {September 2015},
publisher = {VLDB Endowment},
volume = {9},
number = {1},
issn = {2150-8097},
url = {https://doi.org/10.14778/2850469.2850470},
doi = {10.14778/2850469.2850470},
journal = {Proc. VLDB Endow.},
month = sep,
pages = {1–12},
numpages = {12}
}

@online{Microsoft-IndexCompression,
    author = "Microsoft",
    title = "Azure AI Search October Updates: Nearly 100x Compression with Minimal Quality Loss",
    url = "https://techcommunity.microsoft.com/blog/azure-ai-services-blog/azure-ai-search-october-updates-nearly-100x-compression-with-minimal-quality-los/4265447"
}

@online{Mixedbread-BinaryMRL,
    author = "Mixedbread",
    title  = "64 bytes per embedding, yee-haw",
    url    = "https://www.mixedbread.com/blog/binary-mrl"
}

@online{Jina-BinaryEmbeddings,
    author = "Jina",
    title  = "Binary Embeddings: All the AI, 3.125\% of the Fat",
    url    = "https://jina.ai/news/binary-embeddings-all-the-ai-3125-of-the-fat/"
}

@online{Weaviate-BinaryQuantization,
    author = "Weaviate",
    title  = "32x Reduced Memory Usage With Binary Quantization",
    url    = "https://weaviate.io/blog/binary-quantization"
}

@online{MongoDB-BinaryQuantization,
    author = "MongoDB",
    title  = "Binary Quantization \& Rescoring: 96\% Less Memory, Faster Search",
    url    = "https://www.mongodb.com/blog/post/binary-quantization-rescoring-96-less-memory-faster-search"
}

@misc{Thakur2021-arxiv,
  doi = {10.48550/ARXIV.2104.08663},
  author = {Thakur, Nandan and Reimers, Nils and R\"{u}cklé, Andreas and Srivastava, Abhishek and Gurevych, Iryna},
  title = {{BEIR: A Heterogenous Benchmark for Zero-shot Evaluation of Information Retrieval Models}},
  publisher = {arXiv},
  year = {2021},
  copyright = {Creative Commons Attribution Share Alike 4.0 International}
}

@inproceedings{Reimers2021,
    title = "The Curse of Dense Low-Dimensional Information Retrieval for Large Index Sizes",
    author = "Reimers, Nils  and
      Gurevych, Iryna",
    editor = "Zong, Chengqing  and
      Xia, Fei  and
      Li, Wenjie  and
      Navigli, Roberto",
    booktitle = "Proceedings of the 59th Annual Meeting of the Association for Computational Linguistics and the 11th International Joint Conference on Natural Language Processing (Volume 2: Short Papers)",
    month = aug,
    year = "2021",
    address = "Online",
    publisher = "Association for Computational Linguistics",
    url = "https://aclanthology.org/2021.acl-short.77/",
    doi = "10.18653/v1/2021.acl-short.77",
    pages = "605--611"
}

@article{Luan2021,
    title = "Sparse, Dense, and Attentional Representations for Text Retrieval",
    author = "Luan, Yi  and
      Eisenstein, Jacob  and
      Toutanova, Kristina  and
      Collins, Michael",
    editor = "Roark, Brian  and
      Nenkova, Ani",
    journal = "Transactions of the Association for Computational Linguistics",
    volume = "9",
    year = "2021",
    address = "Cambridge, MA",
    publisher = "MIT Press",
    url = "https://aclanthology.org/2021.tacl-1.20/",
    doi = "10.1162/tacl_a_00369",
    pages = "329--345"
}

@inproceedings{Paszke2019,
 author = {Paszke, Adam and Gross, Sam and Massa, Francisco and Lerer, Adam and Bradbury, James and Chanan, Gregory and Killeen, Trevor and Lin, Zeming and Gimelshein, Natalia and Antiga, Luca and Desmaison, Alban and Kopf, Andreas and Yang, Edward and DeVito, Zachary and Raison, Martin and Tejani, Alykhan and Chilamkurthy, Sasank and Steiner, Benoit and Fang, Lu and Bai, Junjie and Chintala, Soumith},
 booktitle = {Advances in Neural Information Processing Systems},
 editor = {H. Wallach and H. Larochelle and A. Beygelzimer and F. d\textquotesingle Alch\'{e}-Buc and E. Fox and R. Garnett},
 pages = {},
 publisher = {Curran Associates, Inc.},
 title = {PyTorch: An Imperative Style, High-Performance Deep Learning Library},
 volume = {32},
 year = {2019}
}

@misc{Sturua2024,
      title={jina-embeddings-v3: Multilingual Embeddings With Task LoRA}, 
      author={Saba Sturua and Isabelle Mohr and Mohammad Kalim Akram and Michael Günther and Bo Wang and Markus Krimmel and Feng Wang and Georgios Mastrapas and Andreas Koukounas and Andreas Koukounas and Nan Wang and Han Xiao},
      year={2024},
      eprint={2409.10173},
      archivePrefix={arXiv},
      primaryClass={cs.CL},
      url={https://arxiv.org/abs/2409.10173}, 
}

@article{Wang2024,
  title={Multilingual E5 Text Embeddings: A Technical Report},
  author={Wang, Liang and Yang, Nan and Huang, Xiaolong and Yang, Linjun and Majumder, Rangan and Wei, Furu},
  journal={arXiv preprint arXiv:2402.05672},
  year={2024}
}

@misc{Merrick2024,
      title={Arctic-Embed: Scalable, Efficient, and Accurate Text Embedding Models}, 
      author={Luke Merrick and Danmei Xu and Gaurav Nuti and Daniel Campos},
      year={2024},
      eprint={2405.05374},
      archivePrefix={arXiv},
      primaryClass={cs.CL},
      url={https://arxiv.org/abs/2405.05374}, 
}

@InProceedings{Fröbe2025,
    author="Fr{\"o}be, Maik and Parry, Andrew and Scells, Harrisen and Wang, Shuai and Zhuang, Shengyao and Zuccon, Guido and Potthast, Martin and Hagen, Matthias",
    editor="Hauff, Claudia and Macdonald, Craig and Jannach, Dietmar and Kazai, Gabriella and Nardini, Franco Maria and Pinelli, Fabio and Silvestri, Fabrizio and Tonellotto, Nicola",
    title="Corpus Subsampling: Estimating the Effectiveness of Neural Retrieval Models on Large Corpora",
    booktitle="Advances in Information Retrieval",
    year="2025",
    publisher="Springer Nature Switzerland",
    address="Cham",
    pages="453--471",
    isbn="978-3-031-88708-6"
}

@misc{Envoldsen2025,
      title={MMTEB: Massive Multilingual Text Embedding Benchmark}, 
      author={Kenneth Enevoldsen and Isaac Chung and Imene Kerboua and Márton Kardos and Ashwin Mathur and David Stap and Jay Gala and Wissam Siblini and Dominik Krzemiński and Genta Indra Winata and Saba Sturua and Saiteja Utpala and Mathieu Ciancone and Marion Schaeffer and Gabriel Sequeira and Diganta Misra and Shreeya Dhakal and Jonathan Rystrøm and Roman Solomatin and Ömer Çağatan and Akash Kundu and Martin Bernstorff and Shitao Xiao and Akshita Sukhlecha and Bhavish Pahwa and Rafał Poświata and Kranthi Kiran GV and Shawon Ashraf and Daniel Auras and Björn Plüster and Jan Philipp Harries and Loïc Magne and Isabelle Mohr and Mariya Hendriksen and Dawei Zhu and Hippolyte Gisserot-Boukhlef and Tom Aarsen and Jan Kostkan and Konrad Wojtasik and Taemin Lee and Marek Šuppa and Crystina Zhang and Roberta Rocca and Mohammed Hamdy and Andrianos Michail and John Yang and Manuel Faysse and Aleksei Vatolin and Nandan Thakur and Manan Dey and Dipam Vasani and Pranjal Chitale and Simone Tedeschi and Nguyen Tai and Artem Snegirev and Michael Günther and Mengzhou Xia and Weijia Shi and Xing Han Lù and Jordan Clive and Gayatri Krishnakumar and Anna Maksimova and Silvan Wehrli and Maria Tikhonova and Henil Panchal and Aleksandr Abramov and Malte Ostendorff and Zheng Liu and Simon Clematide and Lester James Miranda and Alena Fenogenova and Guangyu Song and Ruqiya Bin Safi and Wen-Ding Li and Alessia Borghini and Federico Cassano and Hongjin Su and Jimmy Lin and Howard Yen and Lasse Hansen and Sara Hooker and Chenghao Xiao and Vaibhav Adlakha and Orion Weller and Siva Reddy and Niklas Muennighoff},
      year={2025},
      eprint={2502.13595},
      archivePrefix={arXiv},
      primaryClass={cs.CL},
      url={https://arxiv.org/abs/2502.13595}, 
}

@inproceedings{Cormack2009,
author = {Cormack, Gordon V. and Clarke, Charles L A and Buettcher, Stefan},
title = {{Reciprocal Rank Fusion outperforms Condorcet and Individual Rank Learning Methods}},
year = {2009},
isbn = {9781605584836},
publisher = {Association for Computing Machinery},
address = {New York, NY, USA},
url = {https://doi.org/10.1145/1571941.1572114},
doi = {10.1145/1571941.1572114},
booktitle = {Proceedings of the 32nd International ACM SIGIR Conference on Research and Development in Information Retrieval},
pages = {758–759},
numpages = {2},
location = {Boston, MA, USA},
series = {SIGIR '09}
}

\end{document}